\documentclass[%
 reprint,
 amsmath,amssymb,
 aps,
]{revtex4-2}

\usepackage{graphicx}
\usepackage{dcolumn}
\usepackage{bm}
\usepackage{CJK}
\usepackage[colorlinks,linkcolor=blue,urlcolor=blue,citecolor=blue]{hyperref}

\usepackage{tikz,xcolor,hyperref}

\definecolor{lime}{HTML}{A6CE39}
\DeclareRobustCommand{\orcidicon}{
	\begin{tikzpicture}
	\draw[lime, fill=lime] (0,0) 
	circle [radius=0.16] 
	node[white] {{\fontfamily{qag}\selectfont \tiny ID}};
	\draw[white, fill=white] (-0.0625,0.095) 
	circle [radius=0.007];
	\end{tikzpicture}
	\hspace{-2mm}
}
\foreach \x in {A, ..., Z}{%
	\expandafter\xdef\csname orcid\x\endcsname{\noexpand\href{https://orcid.org/\csname orcidauthor\x\endcsname}{\noexpand\orcidicon}}
}

\begin{document}

\begin{CJK}{UTF8}{gbsn}

\preprint{APS/123-QED}

\title{Hoyle-analog state in $\rm {}^{13}N$}

\author{Ying-Yu Cao (曹颖逾)}
\affiliation{Key Laboratory of Nuclear Physics and Ion-beam Application (MOE), Institute of Modern Physics, Fudan University, Shanghai 200433, China}

\author{De-Ye Tao (陶德晔)}
\affiliation{Key Laboratory of Nuclear Physics and Ion-beam Application (MOE), Institute of Modern Physics, Fudan University, Shanghai 200433, China}

\author{Bo Zhou (周波)\orcidC{}}
\email{zhou\_bo@fudan.edu.cn}
\affiliation{Key Laboratory of Nuclear Physics and Ion-beam Application (MOE), Institute of Modern Physics, Fudan University, Shanghai 200433, China}
\affiliation{Shanghai Research Center for Theoretical Nuclear Physics, NSFC and Fudan University, Shanghai 200438, China}

\author{Yu-Gang Ma (马余刚)\orcidD{}}
\email{mayugang@fudan.edu.cn}
\affiliation{Key Laboratory of Nuclear Physics and Ion-beam Application (MOE), Institute of Modern Physics, Fudan University, Shanghai 200433, China}
\affiliation{Shanghai Research Center for Theoretical Nuclear Physics, NSFC and Fudan University, Shanghai 200438, China}

\date{\today}

\begin{abstract}
We investigate the cluster structure of $\rm {}^{13}N$ using a microscopic $\alpha+\alpha+\alpha+p$ four-body cluster model. The calculated spectra agree well with the observed spectra in the low-lying states. We calculate the reduced width amplitudes and spectroscopic factors to investigate the Hoyle-analog state in $\rm {}^{13}N$. Our calculations show that the $3/2_3^-$ state at $E_x=10.8$ MeV is 
primarily constituted by $\rm {}^{12}C(0^+_2)+\mathit{p}$ and  $\rm {}^{9}B(3/2^-)+\alpha$ components. This finding is generally consistent with the newly observed $3/2^-$ state at $E_x=11.3$ MeV via the $3\alpha + \mathit{p}$ decay channel. Moreover, considering the calculated root-mean-square radius and isoscalar monopole transition strengths, the $3/2_3^-$ state emerges as a candidate for the Hoyle-analog state with the $\rm {}^{12}C(0^+_2)+\mathit{p}$ cluster structure. 
\end{abstract}








\maketitle

\section{Introduction}
\label{introduction}
The Hoyle state of $\rm {}^{12}C$ is characterized by its $3\alpha$ cluster structure~\citep{funaki2003analysis,funaki2005resonance, kanada1998variation,shi2021,zhou2019n,YeYL,MaYG2} with a Bose-Einstein condensate (BEC) nature~\citep{tohsaki2001alpha,ye2024physics}. An interesting topic is the search for the Hoyle-analog state in light nuclei. In self-conjugate $4N$ nuclei, many theoretical works have discussed possible candidates for the Hoyle-analog states in $\rm {}^{16}O$~\citep{ yamada2004dilute, he2014giant, funaki2002description,LiYA,Prasad} and $\rm {}^{20}Ne$~\citep{adachi2021candidates,zhou2012,CaoYT,Ding,Jia}. A study~\citep{funaki2008alpha} based on the orthogonality condition model (OCM) pointed out that the $0_6^+$ state located above the $4\alpha$ threshold is considered a candidate for the $4\alpha$ condensate state. More recently, Tohsaki-Horiuchi-Schuck-R\"opke (THSR)~\citep{funaki2003analysis,funaki2010,zhou2013} calculations predicted that one $5\alpha$ condensate state~\citep{zhou20235} located above the $5\alpha$ threshold, which has a notable amplitude of the $\rm {}^{16}O (0_6^+) + \alpha$ structure.

Another important direction is the search for the Hoyle-analog state in non-self-conjugate nuclei. The $3/2_3^-$ state can be considered as a candidate for the Hoyle-analog state in $\rm {}^{11}B$ and $\rm {}^{11}C$, as investigated by various theoretical models~\citep{nishioka1979structure,descouvemont1996application,kawabata20072alpha+,yamada201082} and experimental measurements~\citep{soic2004,soic2005,Yamaguchi201303,dell2020experimental,li202301}. For example, antisymmetrized molecular dynamics (AMD)~\citep{kanada2007,suhara2012} and generator coordinate method (GCM)~\citep{zhou2018} calculations have shown that the $3/2_3^-$ state of $\rm {}^{11}B$ has a $2\alpha+t$ cluster structure, with a strong isoscalar monopole transition strength from the ground state.

Since $\rm {}^{13}C$ can exhibit the $\rm {}^{12}C + \mathit{n}$ cluster structure, it is important to investigate the possible Hoyle state + neutron structure in the excited states of $\rm {}^{13}C$, and many studies have focused on this aspect.
A previous study~\citep{yamada200621} investigated the spin-orbit splitting in $\rm {}^{13}C$ relative to the Hoyle state of $\rm {}^{12}C$. It was suggested that the reduced density in the Hoyle state leads to a decrease in spin-orbit splitting, particularly in the $1/2^-$ and $3/2^-$ states, which involve the Hoyle configuration and an additional nucleon. OCM calculations~\citep{yamada201592} show that the $1/2_5^+$ state, with a $(0S)^3_{\alpha}(S)_n$ configuration (52\% probability), is a potential candidate for the Hoyle-analog state. Moreover, Ref.~\citep{chiba2020} highlighted that the $1/2_2^+$ state, composed of the $\rm {}^{12}C(0_2^+) \otimes 1\mathit{s}_{1/2}$ configuration, could also be a candidate. More recently, Shin \emph{et al.}~\citep{shin20241} used the real-time evolution method (REM) to propose that the $1/2_4^-$ state might be one candidate for the Hoyle-analog state in $\rm {}^{13}C$.

Building on the studies of $\rm {}^{13}C$, recent experiments have also turned attention to $\rm {}^{13}N$. In particular, $\beta$-delayed charged-particle spectroscopy of $\rm {}^{13}O$~\citep{bishop2024,bishop2023} revealed four new excited states in $\rm {}^{13}N$ at 11.3, 12.4, 13.1, and 13.7 MeV, identified through the $3\alpha + p$ decay channel, with spin-parity assignments of $3/2^-$, $3/2^-$, ($1/2^-$ or $5/2^-$), and $3/2^-$, respectively. Among these, the state at 11.3 MeV ($3/2^-$) is predicted experimentally to might have a mixed configuration of $\rm {}^{9}B(g.s.) \otimes \alpha$ and $\rm {}^{12}C(0_2^+) \otimes \mathit{p}$, making it particularly significant for the search for the Hoyle-analog state. This configuration highlights the need for further theoretical analysis to understand its role in the context of the Hoyle-analog state.

In the present work, we perform GCM calculations to study the cluster structure in $\rm {}^{13}N$, particularly in search of the Hoyle-analog state. We focus on the isoscalar monopole transition strength ($M(IS0)$) and the reduced width amplitudes to discuss the Hoyle-analog states in $\rm {}^{13}N$. 

The paper is organized as follows: In Sec.~\ref{framework}, we provide a concise overview of GCM and the computational details. 
Second, we present the results of the GCM and analyze the cluster structure in Sec.~\ref{results}. Finally, the conclusions of this study are summarized in Sec. ~\ref{summary}.

\section{THEORETICAL FRAMEWORK}
\label{framework}
\subsection{Microscopic cluster wave function and Hamiltonian}
The Hamiltonian is expressed as:
\begin{equation}
\begin{aligned}
\hat{H}=\sum_{i=1}^{13} t_i - t_{\rm{c.m.}}+\sum_{i<j}^{13} V_{ij}^{NN}+\sum_{i<j}^{13} V_{ij}^{C},
\end{aligned}
\end{equation}
where $t_i$  denotes the kinetic energy for the $i$-th nucleon, and $t_{\rm c.m.}$ is the kinetic energy for the center-of-mass. $V^{NN}$ and $V^{C}$ are the effective nucleon-nucleon interaction and Coulomb interaction, respectively. The effective nucleon-nucleon interaction includes the Volkov No.2~\citep{volkov1965equilibrium} and the spin-orbit part of the G3RS interaction~\citep{yamaguchi1979effective,okabe1979structure}, which is given as
\begin{equation}
\begin{aligned}
V_{ij}^{NN}=&\sum_{n=1}^2 V_ne^{-r_{ij}^2/a_n^2}(W+BP_{\sigma}-HP_{\tau}-MP_{\sigma}P_{\tau}) \\
&+\sum_{n=1}^2 w_n e^{-b_nr_{ij}^2} P({}^{3}O)\boldsymbol{L} \cdot \boldsymbol{S}, 
\end{aligned}
\end{equation}
where $P({}^{3}O)$ represents the projection operator to the triplet-odd states,  $V_1 = - 60.65$ MeV, $V_2$ = 61.14 MeV, $a_1$ = 1.80 fm, $a_2$ = 1.01 fm, $W = 1 - M$, $M$ = 0.60, and $B=H=0.125$. For the G3RS spin-orbit term, $b_1 = 5.0 \rm \ fm^{-2}$, $b_2 = 2.778 \rm \ fm^{-2}$, and $w_1=-w_2$ = 2000 MeV.

\begin{figure}[htbp!]
  \centering  \includegraphics[width=7.5cm]{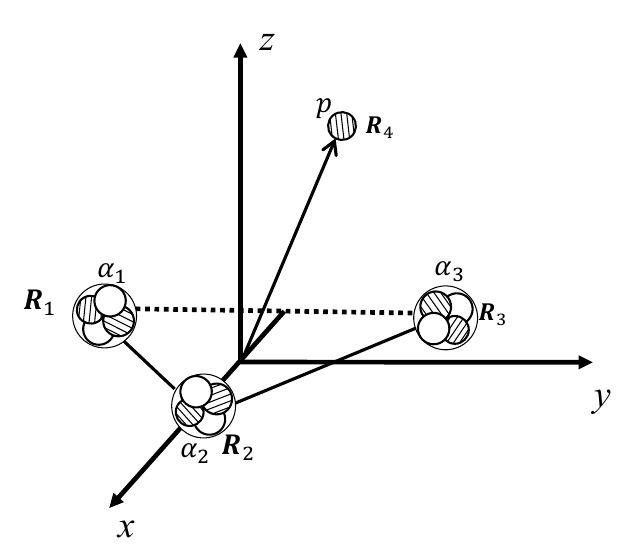}
  \caption{ Schematic diagram of $\alpha+\alpha+\alpha+p$ clustering structure of the Brink wave function of $\rm{}^{13}N$.}
  \label{13N_configration}
\end{figure}

The Brink wave function~\citep{brink1965alpha} of $\rm {}^{13}N$ is constituted by the $ \alpha+\alpha+\alpha+p$ cluster configuration as shown in Fig~\ref{13N_configration}, 

\begin{equation}
\begin{aligned}
&\Phi(\boldsymbol{R}_1,\boldsymbol{R}_2,\boldsymbol{R}_3,\boldsymbol{R}_4)=\mathcal{A}\{\Phi_{\alpha}(\boldsymbol{R}_1)\Phi_{\alpha}(\boldsymbol{R}_2)\Phi_{\alpha}(\boldsymbol{R}_3)\Phi_{p}(\boldsymbol{R}_4)\},\\
\end{aligned}
\end{equation}
where $\Phi_{\alpha}$ is the wave function of $\alpha$ cluster and denote as 
\begin{equation}
\begin{aligned}
&\Phi_{\alpha}(\boldsymbol{R})=\mathcal{A}\{\prod_{i=1}^{4}\phi(\boldsymbol{R},\boldsymbol{r}_i)\chi_i \tau_i\}.\\
\end{aligned}
\end{equation}
$\boldsymbol{R}_1$, $\boldsymbol{R}_2$, $\boldsymbol{R}_3$, and $\boldsymbol{R}_4$ are the generator coordinates of three $\alpha$ clusters and valence proton, abbreviated as $\{\boldsymbol{R}\}=\{\boldsymbol{R}_1, \boldsymbol{R}_2, \boldsymbol{R}_3, \boldsymbol{R}_4 \}$ and the condition
$(4\boldsymbol{R}_1+4\boldsymbol{R}_2+4\boldsymbol{R}_3+\boldsymbol{R}_4)/13=0$ is applied. $\phi(\boldsymbol{R},\boldsymbol{r}_i)\chi_i \tau_i$ represents the $i$-th single-particle wave function, where $\phi(\boldsymbol{R},\boldsymbol{r}_i)$ is the spatial wave function and denote as 
\begin{equation}
\begin{aligned}
&\phi(\boldsymbol{R},\boldsymbol{r}_i)=(\frac{1}{\pi b^2})^{3/4}e^{-\frac{(\boldsymbol{r}_i-\boldsymbol{R})^2}{2b^2}},
\end{aligned}
\end{equation}
$\chi_i$ and $\tau_i$ denote the spin and isospin of each nucleon. The harmonic-oscillator parameter is set to $b$=1.46 fm.

The GCM wave functions of $\rm {}^{13}N$ are obtained by superposing over various $ \alpha+\alpha+\alpha+p$ configurations, which can be written as 
\begin{equation}
\begin{aligned}
\Psi_M^{J \pi}=\sum_{\{\boldsymbol{R}\}K}f_{\{\boldsymbol{R}\}K} \Phi_{MK}^{J \pi}(\{\boldsymbol{R}\}),
\end{aligned}
\end{equation}
where $\Phi_{MK}^{J \pi}(\{\boldsymbol{R}\})$ is the projected wave function,
\begin{equation}
\begin{aligned}
\Phi_{MK}^{J \pi}(\{\boldsymbol{R}\})=P_{MK}^JP^{\pi}\Phi(\{\boldsymbol{R}\}).
\end{aligned}
\end{equation}
$P_{MK}^J$ and $P^{\pi}$ denote the parity and angular momentum projector, respectively. The coefficients of the superposition $f_{\{\boldsymbol{R}\} K}$ are determined by solving the Hill-Wheeler equation~\cite{ring2004nuclear}, which is given by,
\begin{equation}
\begin{aligned}
\sum_{\{\boldsymbol{R}'\}K'}&f_{\{\boldsymbol{R}'\}K'} \bigg[ \langle \Phi_{MK}^{J \pi}(\{\boldsymbol{R}\})
|\hat{H}|
\Phi_{MK'}^{J \pi}(\{\boldsymbol{R}'\}) \rangle -\\
&E\langle \Phi_{MK}^{J \pi}(\{\boldsymbol{R}\})                      
|\Phi_{MK'}^{J \pi}(\{\boldsymbol{R}'\}) \rangle
\bigg]=0.
\end{aligned}
\end{equation}
 
\subsection{Reduced width amplitudes and spectroscopic factors}
To investigate configurations of clustering states in $\rm {}^{13}N$, we calculate the reduced width amplitudes (RWAs) for the $\rm {}^{12}C + \mathit{p}$ and $\rm {}^{9}B + \alpha$ channels. The RWA in the two-body channel is defined as 
\begin{equation}
\begin{aligned}
y_{j_1 \pi_1 j_2 \pi_2 j_{12}l}^{J{\pi}}&(a)=\sqrt{\frac{A!}{\left(1+\delta_{C_1 C_2}\right) C_{1}!C_{2}!}} \times \\
&\left\langle\left.\frac{\delta(r-a)}{r^2}\left[Y_l(\hat{r})\left[\Phi_{C_1}^{j_1 \pi_1} \Phi_{C_2}^{j_2 \pi_2}\right]_{j_{12}}\right]_{J M} \right\rvert\, \Psi_{M}^{J \pi}\right\rangle,
\label{rwa}
\end{aligned}
\end{equation}
where $\Phi_{C_1}^{j_1 \pi_1}$ and $\Phi_{C_2}^{j_2 \pi_2}$ are the wave functions of cluster $C_1$ and $C_2$. The spins $j_1$ of cluster $C_1$ and $j_2$ of cluster $C_2$ are coupled to $j_{12}$, then the $j_{12}$ is coupled to the orbital angular momentum $l$ of the relative-motion function to yield the total spin-parity $J^{\pi}$. The parity $\pi_1$, $\pi_2$ and the orbital angular momentum $l$ satisfy the relation $\pi=\pi_1\pi_2(-)^{l}$. 
The spectroscopic factor of the two-body channel is defined as the integrals of the RWAs,
\begin{equation}
\begin{aligned}
S^{2}_{j_1 \pi_1 j_2 \pi_2 j_{12}l}=\int_0^{\infty} d a\left|a y_{j_1 \pi_1 j_2 \pi_2 j_{12} l}^{J \pi }(a)\right|^2
\label{rwa}
\end{aligned}
\end{equation}

To evaluate the RWAs, we use the Laplace expansion method proposed in Ref.~\citep{chiba2017} and more details of RWAs can be found in Ref.~\cite{lyu2018,lyu2019,tao2024reduced}.

\section{RESULTS AND DISCUSSIONS}
\label{results}
We perform GCM calculations by superposing 1000 basis wave functions. The generated coordinates for the three $\alpha$ clusters ($\alpha_1$, $\alpha_2$, $\alpha_3$) and the proton ($p$) are randomly generated according to the normal distribution.

\begin{figure*}[]
  \centering
\includegraphics[width=0.7\textwidth]{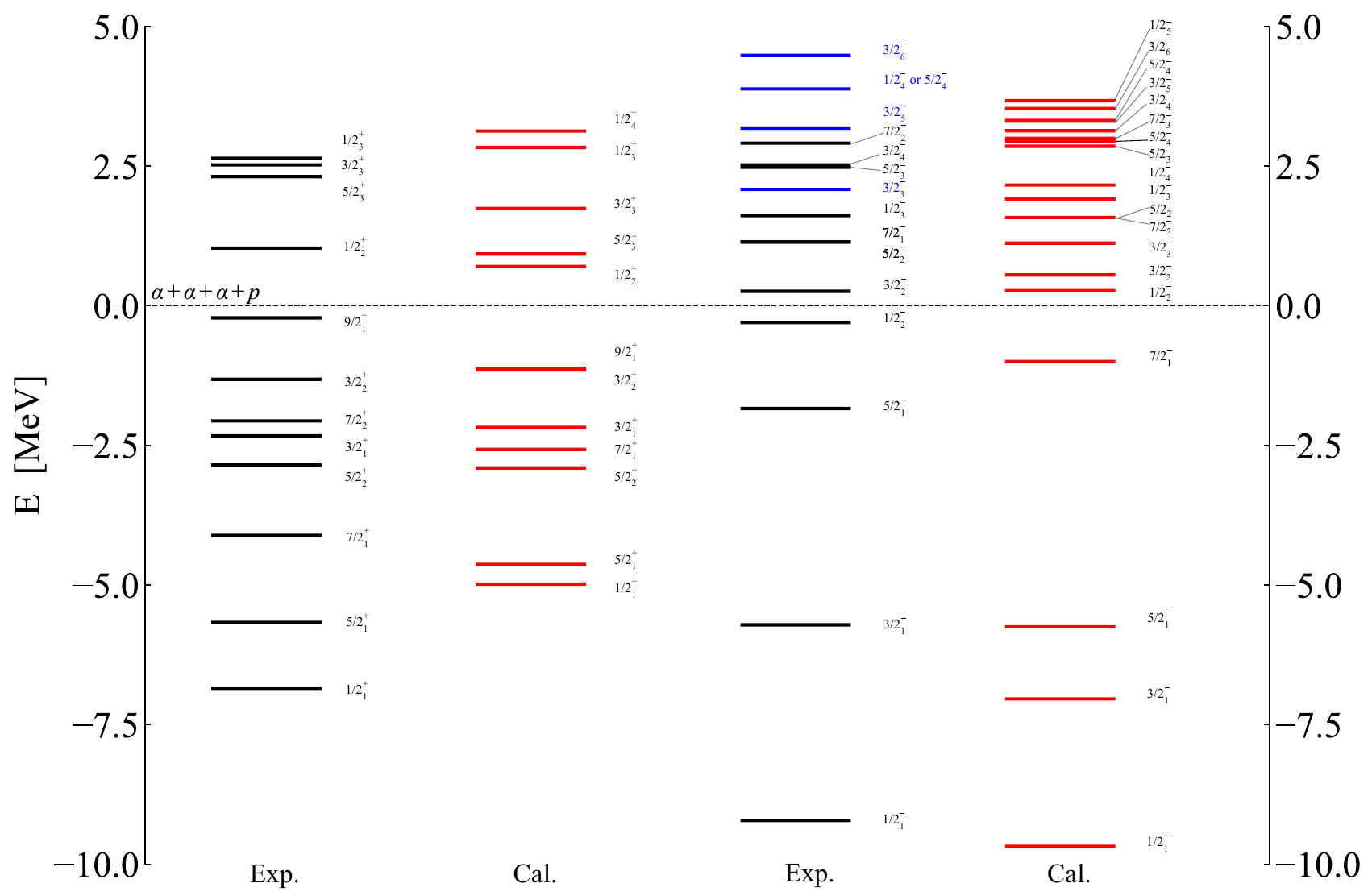}
  \caption{(Color online) Present calculation and experimental energy
spectra of $\rm {}^{13}N$. States marked in blue in the experimental spectra are newly observed in the recent experiment~\cite{bishop2024}. The dashed line represents the 3$\alpha+p$ threshold.}
  \label{Elevel}
\end{figure*}

Figure~\ref{Elevel} shows the energy spectra obtained from the GCM calculations, compared to the observed experimental data. The blue-marked states in the experimental spectra are newly observed in the recent experiment~\cite{bishop2024, bishop2023}. 
It should be noted that there remain significant discrepancies between the calculated and experimental values for the low-lying states, specifically those below the 3$\alpha+p$ threshold. These discrepancies may be attributed to the limitations in modeling the interactions, particularly the inadequate consideration of spin-orbit interactions. Additionally, we compare theoretical calculations with experimental values for states within 5 MeV above the threshold. This region is a key focus in our search for the Hoyle state plus proton structure.


The calculated electric dipole transition strengths ($B(E1)$) for low-lying states are presented in Table~\ref{transition}, alongside comparisons with experimental data. The calculated value of $B(E1, 1/2^+_1 \to 1/2^-_1) = 0.0007$ $e^2~\rm fm^2$ was notably underestimated. Furthermore, the predicted quadrupole transition strengths ($B(E2)$) are also shown in Table~\ref{transition}.


\begin{table}[htbp]
\caption{The electric dipole transition strengths $B(E1)$ (units:~$ e^2~\rm fm^2$) and quadrupole transition strengths $B(E2)$(units:~$ e^2~\rm fm^4$) obtained by the present study and experiment~\cite{ajzen1991}.}
\label{transition}
\begin{ruledtabular}
\begin{tabular}{ccc}
Transition &
Present &
Experiment   \\  \hline
$B(E1,3/2_1^- \to 1/2_1^+)$ &
0.016 & 
0.036 \\ 
$B(E1,1/2_1^+ \to 1/2_1^-)$ &
0.0007 & 
0.036 $\pm$ 0.004 
\\ 
$B(E2,3/2_1^- \to 1/2_1^-)$ &
4.87 & 
 \\ 
$B(E2,5/2_1^- \to 1/2_1^-)$ &
3.71 & 
 \\ 
\end{tabular}
\end{ruledtabular}
\end{table}

\begin{figure}[h]
  \centering
\includegraphics[width=0.55\textwidth]{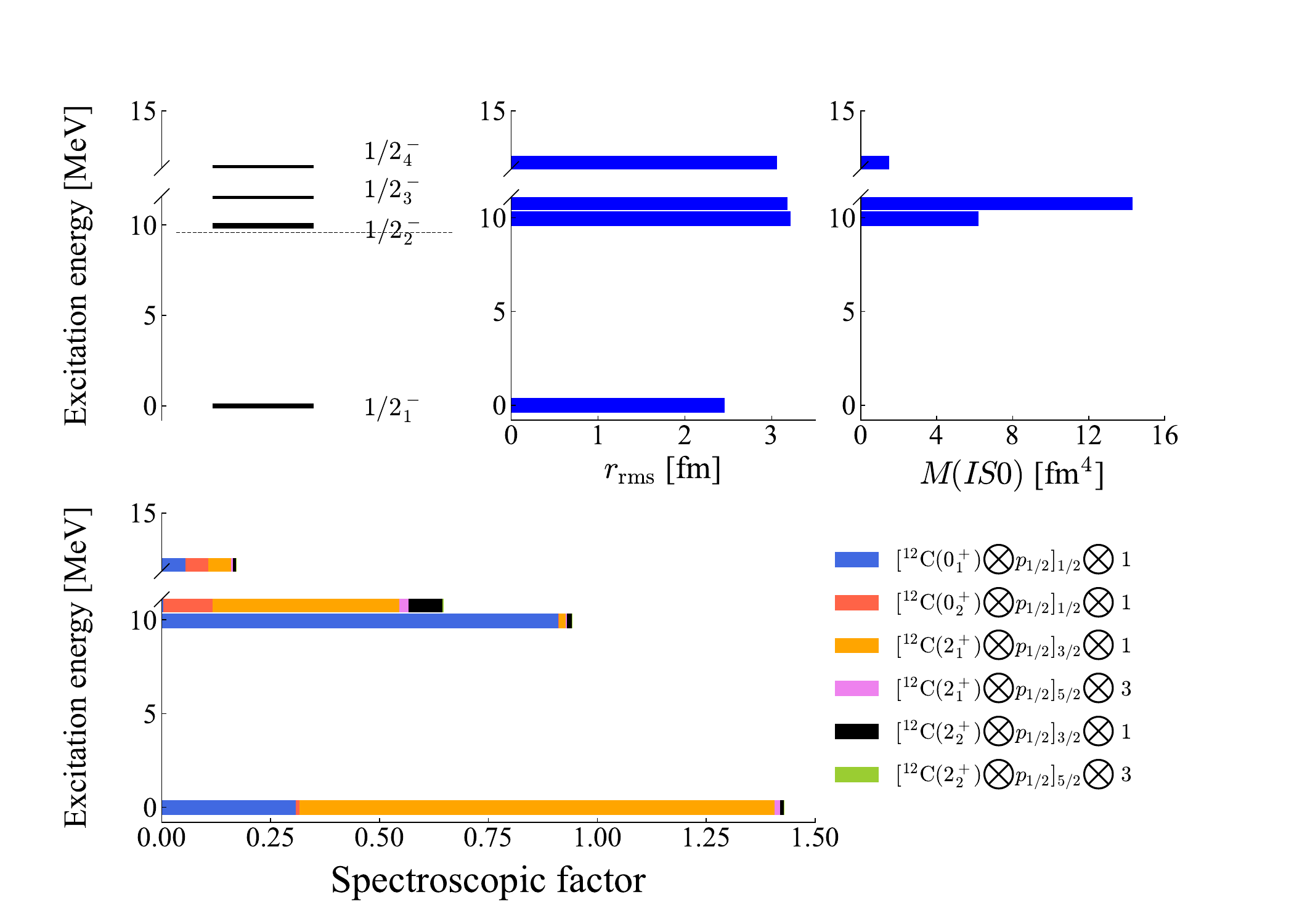}
  \caption{(Color online) The excitation energies and the properties of the $1/2^-$ states obtained in the present work. 
  The excitation energies, $r_{\rm rms}$, and $M(IS0)$ from the excited states to the ground state are presented in the upper panel. The calculated spectroscopic factors ($S^2$) in the $\rm {}^{12}C+\mathit{p}$ channels are presented in the lower panel. The dashed line represents the 3$\alpha+p$ threshold.}
  \label{neg0.5properties}
\end{figure}

Next we discuss the structure of the $1/2^-$ states in $\rm {}^{13}N$. The calculated root-mean-square (r.m.s.) radii, spectroscopic factors for the $\rm {}^{12}C+\mathit{p}$ channels, and isoscalar monopole transition strengths $M(IS0)$ are presented in Fig.~\ref{neg0.5properties}.

As shown in Fig.~\ref{neg0.5properties}, the ground state exhibits a compact shell-model structure with an r.m.s.\ radius of 2.46 fm. This shell-model-like nature is supported by the large spectroscopic factors in the $[\rm {}^{12}C(2_1^+)\bigotimes \mathit{p}_{1/2}]_{3/2} \bigotimes 1$ and $[\rm {}^{12}C(0_1^+)\bigotimes \mathit{p}_{1/2}]_{1/2} \bigotimes 1$ channels, which are 1.09 and 0.31, respectively.
In contrast to the ground state, the excited $1/2^-$ states exhibit r.m.s.\ radii exceeding 3.0 fm, indicating a possible development of cluster structures. Notably, the $1/2_2^-$ state displays a large RWA amplitude in the $[\rm {}^{12}C(0_1^+)\bigotimes \mathit{p}_{1/2}]_{1/2} \bigotimes 1$ channel ($S^2$ = 0.91) with one node, suggesting that it represents a nodal excitation of the inter-cluster motion between $\rm {}^{12}C~(0_1^+)$ and the proton. Therefore, the $1/2_2^-$ state is identified as an $\rm{}^{12}C(0_1^+)+\mathit{p}$ cluster state rather than a Hoyle-analog state.
The $1/2_3^-$ state exhibits a significant isoscalar monopole transition strength, while the RWAs indicate that both the $1/2_3^-$ and $1/2_4^-$ states have minimal contributions to the $[\rm {}^{12}C(0_2^+)\bigotimes \mathit{p}_{1/2}]_{1/2} \bigotimes 1$ channel. Consequently, neither the $1/2_3^-$ nor the $1/2_4^-$ state can be considered as Hoyle-analog states.



Next, we discuss cluster structure of the $1/2^+$ states.
The r.m.s.\ radius of the $1/2_1^+$ state is 2.73 fm, which is slightly large than that of ground state. The spectroscopic factor in the $[
\rm {}^{12}C(0_1^+)\bigotimes \mathit{p}_{1/2}]_{1/2}$ channel is 0.84, with small contributions from other channels. These features suggest that the $1/2_1^+$ state has a compact shell structure. In contrast, the $1/2_2^+$ state shows an r.m.s.\ radius of 3.19 fm, suggesting a more developed cluster structure. The RWAs in the $[\rm {}^{12}C(0_1^+)\bigotimes \mathit{p}_{1/2}]_{1/2} \bigotimes 1$ channel dominate, with a spectroscopic factor of $S^2 = 0.69$. The RWAs in this channel exhibit two nodes, which suggests that the $1/2_2^+$ state can be interpreted as a nodal excitation of the inter-cluster motion between $\rm {}^{12}C$ and the valence proton. The $1/2_3^+$ state, on the other hand, has spectroscopic factors of less than 0.2 in all channels, even though both the r.m.s.\ radius and isoscalar monopole transition strength ($M(IS0)$) show noticeable enhancements. This indicates a more diffuse structure with no strong dominance in any specific cluster configuration. Similarly, the $1/2_4^+$ state displays small RWAs in the $[\rm {}^{12}C(0_2^+)\bigotimes \mathit{p}{1/2}]{1/2} \bigotimes 1$ channel, with a spectroscopic factor magnitude lower than 0.04.
Based on these observations, we conclude that none of the $1/2^+$ states can be identified as a Hoyle-analog state. 



\begin{figure}[hbpt!]
  \centering
\includegraphics[width=0.55\textwidth]{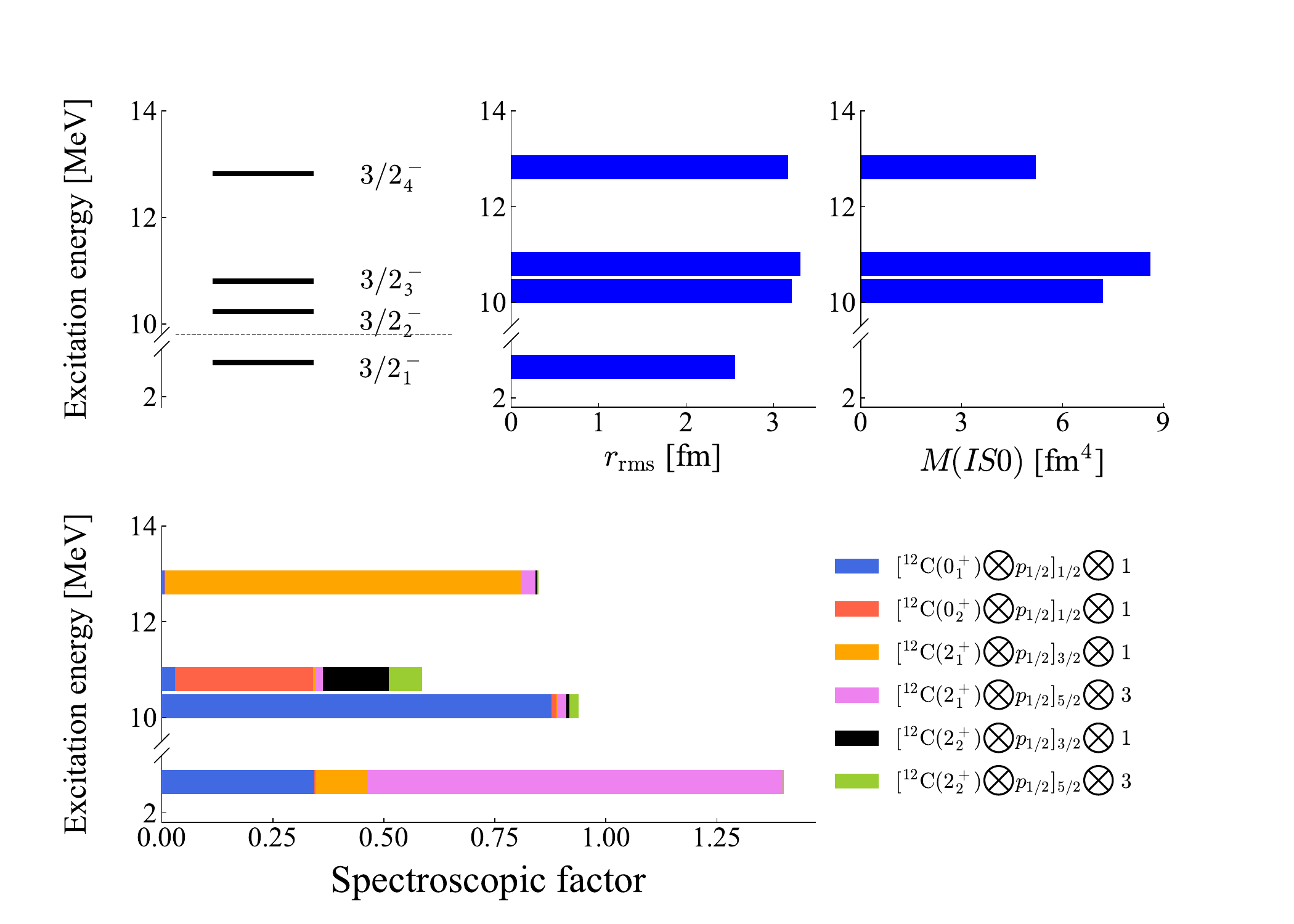}
  \caption{(Color online) The excitation energies and the properties of the $3/2^-$ states obtained in the present work. The excitation energies, $r_{\rm rms}$, and $M(IS0)$ from the excited states to the $3/2_1^-$ state are presented in the upper panel. The calculated spectroscopic factors ($S^2$) in the $\rm {}^{12}C+\mathit{p}$ channel are presented in the lower panel. The dashed line represents the 3$\alpha+p$ threshold.}
  \label{neg1.5}
\end{figure}

The properties of the $3/2^-$ states are summarized in Fig.~\ref{neg1.5}. The calculated r.m.s.\ radius of the $3/2^-_1$ state is 2.56 fm, suggesting it has a compact shell structure. This compact structure is further supported by large spectroscopic factors in the $[
\rm {}^{12}C(2_1^+)\bigotimes \mathit{p}_{1/2}]_{5/2} \bigotimes 1$ and $[
\rm {}^{12}C(0_1^+)\bigotimes \mathit{p}_{1/2}]_{1/2} \bigotimes 1$ channels, with magnitudes of 0.93 and 0.34 respectively. 

\begin{figure}[h]
  \centering
\includegraphics[width=8.0cm]{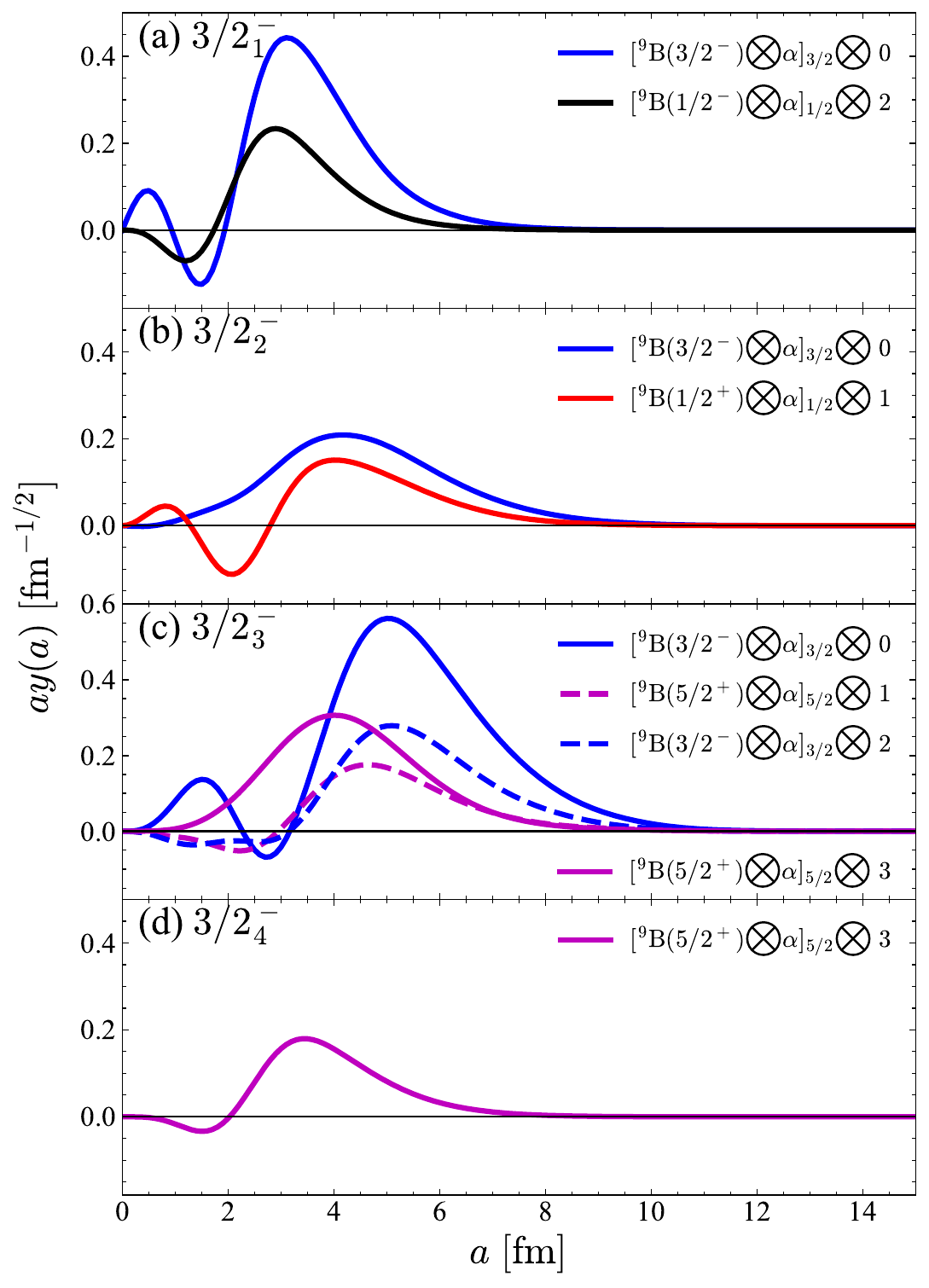}
  \caption{(Color online) The calculated RWAs of the $3/2^-$ states in the $\rm {}^{9}B+\mathit{\alpha}$ channels. The RWAs that yield spectroscopic factor large than 0.04 are displayed.}
  \label{neg1.5rwa_alpha_9B}
\end{figure}

The $3/2_2^-$ state, located 0.26 MeV above the $3\alpha+\mathit{p}$ threshold, has its largest spectroscopic factor of 0.88 in the $[\rm {}^{12}C(0_1^+)\bigotimes \mathit{p}_{1/2}]_{1/2} \bigotimes 1$ channel. The RWAs for this channel display one node, similar to that of the $1/2_2^-$ state. Moreover, the $[\rm {}^{9}B(3/2^-) \bigotimes \alpha]_{3/2} \bigotimes 0$ channel contributes modestly to the $3/2_2^-$ state, with a spectroscopic factor of 0.12, as shown in Fig.~\ref{neg1.5rwa_alpha_9B}(b).

\begin{figure}[hbp]
  \centering
\includegraphics[width=8.0cm]{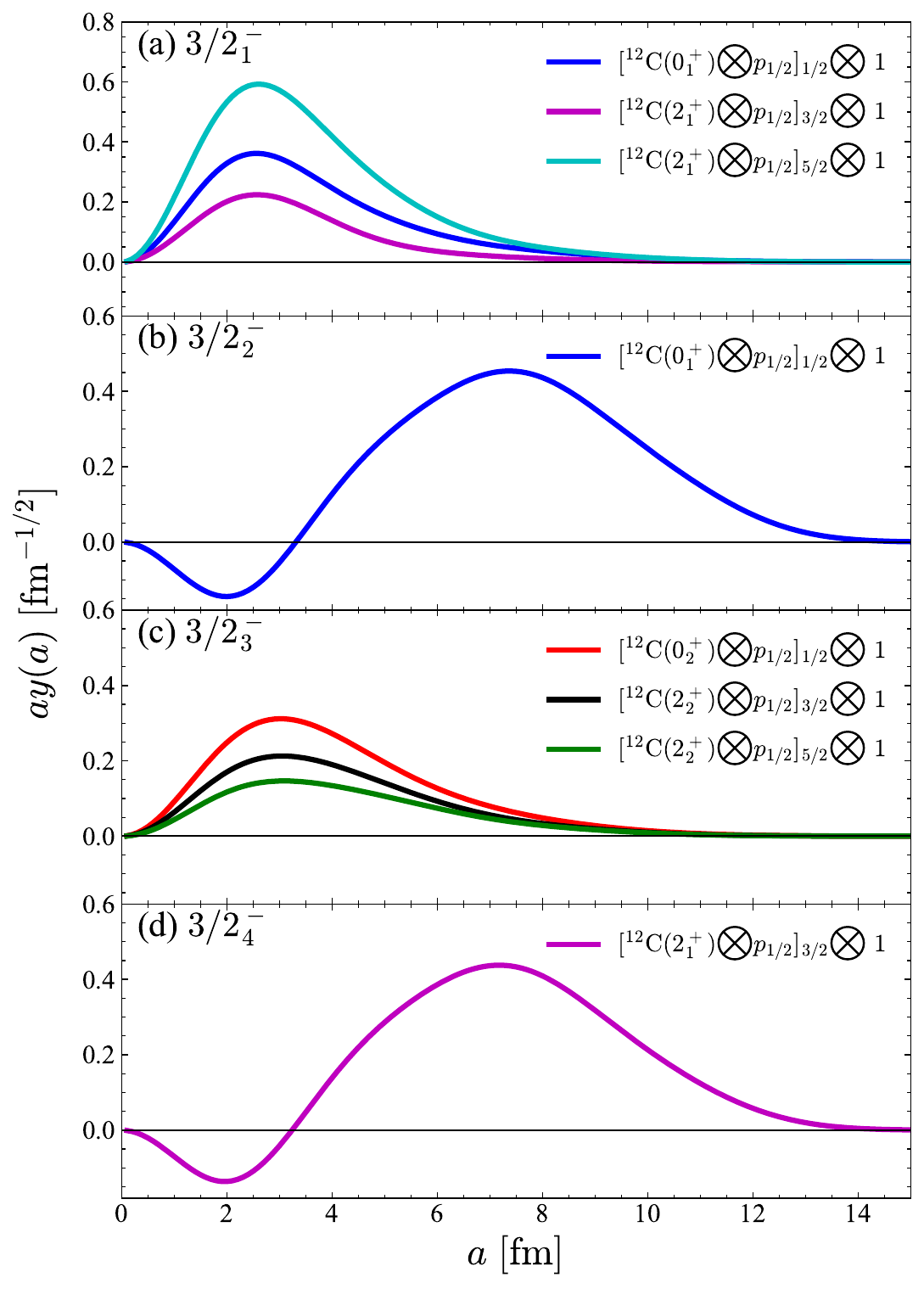}
  \caption{(Color online) The calculated RWAs of the $3/2^-$ states in the $\rm {}^{12}C+\mathit{p}$ channels. The RWAs that yield spectroscopic factor large than 0.04 are displayed.}
  \label{neg1.5rwa}
\end{figure}

The $3/2_3^-$ state at $E_x=10.8$ MeV is close to the newly observed state at $E_x=11.3$ MeV~\cite{bishop2024}. As shown in Fig.~\ref{neg1.5rwa}(c) and Fig.~\ref{neg1.5rwa_alpha_9B}(c), the calculated RWAs in the $[
\rm {}^{12}C(0_2^+)\bigotimes \mathit{p}_{1/2}]_{1/2} \bigotimes 1$ ($S^2$=0.31) and $[\rm {}^{9}B(3/2^-) \bigotimes \alpha]_{3/2} \bigotimes 0$ ($S^2$=0.73) channels are dominant, which is generally consistent with the experimental observations~\cite{bishop2024}. On the other hand, the $3/2_3^-$ state has a non-negligible magnitude of spectroscopic factor ($S^2$=0.15) in the $[
\rm {}^{12}C(2_2^+)\bigotimes \mathit{p}_{1/2}]_{3/2} \bigotimes 1$ channel. The second $2^+$ state is considered to be related to the Hoyle state, as a member of the Hoyle band, and exhibits characteristics of a dilute gas-like state \citep{ugeaki1977,zimmerman2013,itoh2011c,funaki2015h}. The $\rm {}^{12}C(0_2^+)\bigotimes \mathit{p}$ and $\rm {}^{12}C(2_2^+)\bigotimes \mathit{p}$ configurations, combined with an enhanced radius, significant monopole transition strength, and the RWA contributions in both the $[\rm {}^{12}C(0_2^+)\bigotimes \mathit{p}_{1/2}]_{1/2} \bigotimes 1$ and $[\rm {}^{12}C(2_2^+)\bigotimes \mathit{p}_{1/2}]_{3/2} \bigotimes 1$ channels, suggest that the $3/2_3^-$ state can be regarded as a candidate for the Hoyle-analog state.

To investigate the dilute gas-like nature of the $3/2_3^-$ state, we calculate its GCM-Brink overlap . The maximum overlap is found to be 0.41, corresponding to inter-cluster distances of $D_{\alpha_1-\alpha_2}=2.77$ fm,  $D_{\alpha_1-\alpha_3}=4.29$ fm, $D_{\alpha_2-\alpha_3}=6.14$ fm, and a distance of $D_{3\alpha-p}=0.53$ fm between the center of mass of the three $\alpha$ and the proton. In comparison, the ground state shows a maximum overlap of 0.91, with more compact inter-cluster distances: $D_{\alpha_1-\alpha_2}=2.54$ fm, $D_{\alpha_1-\alpha_3}=2.44$ fm, $D_{\alpha_2-\alpha_3}=2.42$ fm, and $D_{3\alpha-p}=2.52$ fm. 
The larger spatial distribution observed in the $3/2_3^-$ state, as revealed by the overlap function, supports its dilute gas-like cluster structure.

The $3/2_4^-$ state is located at an excitation energy of 12.6 MeV. The RWAs presented in Fig.~\ref{neg1.5rwa}(d). It is noted that the $[
\rm {}^{12}C(2_1^+)\bigotimes \mathit{p}_{1/2}]_{3/2} \bigotimes 1$ channel is dominant ($S^2=0.80$), while other channels provide minor contributions. Additionally, the $[\rm {}^{9}B(5/2^+) \bigotimes \alpha]_{5/2}\bigotimes 3$ channel has a small magnitude of spectroscopic factor ($S^2=0.06$) as shown in Fig.~\ref{neg1.5rwa_alpha_9B}(d). According to Ref.~\cite{bishop2024}, experimentally observed $3/2^-$ states at excitation energies of 11.8 and 12.4 MeV are associated with $\rm {}^{12}C(0_1^+)+\mathit{p}$ and $\rm {}^{9}B(1/2^+)\bigotimes \alpha$ configurations, respectively. The channel contributions obtained from our calculated RWAs differ from these experimental observations.

\section{Summary and conclusions}
\label{summary}
In this study, we investigate the cluster structure of $\rm {}^{13}N$ using the GCM. Our calculations basically reproduce the energy spectra of the low-lying states for both negative-parity and positive-parity states.
We analyze the r.m.s.\ radii, isoscalar monopole transition strengths, and RWAs for the $1/2^-$, $3/2^-$, and $1/2^+$ states. 

For the $1/2_2^-$, $3/2_2^-$, and $1/2_2^+$ states, the GCM-Brink overlap calculations reveal a compact $3\alpha$ spatial configuration, with a significant separation between the $p$ and the $3\alpha$. The RWAs indicate that these states are predominantly characterized by the $[
\rm {}^{12}C(0_1^+)\bigotimes \mathit{p}_{1/2}]_{1/2}$ configuration, with nodal excitation of the inter-cluster motion between $\rm {}^{12}C$ and $p$. Therefore, we conclude that these states possess a two-body $\rm {}^{12}C+\mathit{p}$ cluster structure. On the other hand, the $1/2_3^-$, $1/2_4^-$, $1/2_3^+$, and $1/2_4^+$ states have non-negligible spectroscopic factors in the $\rm {}^{12}C(0_2^+)+\mathit{p}$ channel but their magnitudes remain small. As a result, these states cannot be considered as Hoyle-analog states.

 The $3/2_3^-$ state, located at 10.8 MeV, has a significant r.m.s.\ radius and enhanced isoscalar monopole transition strength. The overlap calculation shows a broad spatial distribution of the $3\alpha$ cluster with valence proton positioned close to the $3\alpha$ cluster, suggesting a dilute gas-like nature. Moreover, The RWAs for the $3/2_3^-$ state reveal a notable contribution from the $[
\rm {}^{12}C(0_2^+)\bigotimes \mathit{p}_{1/2}]_{1/2} \bigotimes 1$ channel with $S^2=0.31$ and a non-negligible component in the $[
\rm {}^{12}C(2_2^+)\bigotimes \mathit{p}_{1/2}]_{3/2} \bigotimes 1$ channel. Therefore, we propose that the $3/2_3^-$ state is a candidate for the Hoyle-analog state.

\section*{Acknowledgements}
This work is supported by the National Key R$\&$D Program of China (2023YFA1606701). This work was supported in part by the National Natural Science Foundation of China under contract Nos. 12175042, 11890710, 11890714, 12047514, and 12147101, Guangdong Major Project of Basic and Applied Basic Research No. 2020B0301030008, and China National Key R$\&$D Program No. 2022YFA1602402. This work was partially supported by the 111 Project.











%

\end{CJK}
\end{document}